\documentclass[aps,showpacs,amsmath,amssymbd,dvips,showkeys,preprint,preprintnumbers]{revtex4}

\usepackage{graphicx}
\usepackage{color}

\def \be  {\begin{equation}}
\def \ee  {\end{equation}}
\def \ee  {\end{equation}}
\def \bea {\begin{eqnarray}}
\def \eea {\end{eqnarray}}

\begin{document}

\preprint{ECTP-2016-11}
\preprint{WLCAPP-2016-11}
\hspace{0.05cm}
\title{SU($3$) Polyakov linear-sigma model: Magnetic properties of QCD matter in thermal and dense medium}

\author{Abdel Nasser  Tawfik}
\email{a.tawfik@eng.mti.edu.eg}
\affiliation{Egyptian Center for Theoretical Physics (ECTP), Modern University for Technology and Information (MTI), 11571 Cairo, Egypt}
\affiliation{World Laboratory for Cosmology And Particle Physics (WLCAPP), 11571 Cairo, Egypt}

\author{Abdel Magied Diab}
\email{a.diab@eng.mti.edu.eg}
\affiliation{Egyptian Center for Theoretical Physics (ECTP), Modern University for Technology and Information (MTI), 11571 Cairo, Egypt}
\affiliation{World Laboratory for Cosmology And Particle Physics (WLCAPP), 11571 Cairo, Egypt}

\date{\today}

\begin{abstract}
The linear-sigma model, in which information about confining gluons is included through the Polyakov-loop potential (PLSM), is considered in order to perform a systematic study for various magnetic properties of QCD matter under extreme conditions of high temperatures and densities and finite magnetic field strengths. The introduction of magnetic field to the PLSM Lagrangian requires suitable utilization of Landau quantization, modification in the dispersion relations, and momentum-space dimension-reduction. We observed that increasing the magnetic field leads to filling-up lower Landau levels first and decreasing the number of occupied levels. We conclude that the population of Landau levels is most sensitive to the magnetic field and to the quark charges. The influences of finite magnetic field on the temperature dependence of chiral and deconfinement order-parameter(s) are studied. We present estimations for the magnetization, the magnetic susceptibility, the permeability and the catalysis properties of QCD matter as functions of temperature. The dependences of the resulting freezeout parameters, temperatures and baryon chemical potentials on the corresponding magnetic field strengths have been analyzed, as well. These calculations are compared with recent lattice QCD simulations, whenever available. We conclude that the QCD matter seems to have paramagnetic property at temperatures greater than the critical one. There is an evidence for weak diamagnetic property at low temperatures. Last but not least, we observe that the magnetic catalysis is inverse, namely the critical temperatures decrease with increasing the magnetic field.  
\end{abstract}

\pacs{11.10.Wx, 25.75.Nq, 98.62.En, 12.38.Cy}
\keywords{Chiral transition, magnetic fields, magnetic catalysis, critical temperature}

\maketitle


\section{Introduction \label{intro}}

The systematic study of strongly interacting QCD matter under extreme conditions of high temperatures and densities and finite magnetic fields belongs to the ultimate goals of the heavy-ion collision (HIC) facilities such as Superproton Synchrotron (SPS) at CERN, Relativistic Heavy-Ion Collider (RHIC) at BNL, the Large Hadron Collider (LHC) at CERN, and the future facilities such as the Nuclotron-based Ion Collider FAcility (NICA) at JINR and the Facility for Antiproton and Ion Research (FAIR) at GSI. It is conjectured that in HIC under such extreme conditions, chiral and deconfinement order-parameters from hadron to quark-gluon plasma (QGP) likely take place. The chiral structure of hadrons, the properties of QGP and the location of the critical endpoint (CEP) in the phase diagram are examples on significant researches performed over the last decades.  

In the present work, we address the temperature dependence of the corresponding order-parameters at finite magnetic field strengths. We utilize the extended SU($3$) linear-sigma model to study different magnetic properties such as  magnetization, magnetic susceptibility and permeability. Moreover, we estimate the chiral phase-diagram; temperature vs. baryon chemical potential in varying magnetic field strengths.  

Due to oppositely directed relativistic motion of charges especially in off-central collisions, a huge magnetic field can be created in HIC. Because of the very short lifetime of such a magnetic field, it is assumed that the generated field has almost no effect on the detector and on its external magnet but a remarkable influence on the strongly interacting QCD matter. The expected magnetic field at LHC, RHIC and SPS energies ranges from $10-15 \, m_{\pi}^2$,  $m_{\pi}^2$ down to $0.1\, m_{\pi}^2$, respectively \cite{Skokov:2009, Elec:Magnet}, where $m_{\pi}^2$, the pion mass squared,  is equivalent to $\sim 10^8$ Gauss.

The largest effect of such magnetic fields causes not only catalysis in the chiral symmetry breaking, i.e. decreasing the critical temperature with increasing magnetic field \cite{Bruckmann:2013,Preis:2011A,Preis:2011B}, but also in the chiral phase-structure of the produced hadrons. It causes modifications in the nature of the chiral phase-transition, as well \cite{Sanfilippo:2010, catalysis:2014, Catalysis:2015} and in the energy loss due to the synchrotron radiation of the quarks \cite{Synchrotron:2010, Elec:Magnet}. These effects are not necessarily limited to the early stages of HIC. During later stages, the response of QCD matter to finite magnetic field is assumed to have a large medium-dependence, which obviously depends on the diffusion time variation \cite{Synchrotron:2010, Elec:Magnet} and the electric conductivity of the medium \cite{Gupta:2004, Bratkovskaya:1995}. 

On the other hand, characterizing the magnetic field effects is closely related to essential properties of the QCD matter such as the chiral magnetic effect (CME) and the magnetic catalysis \cite{Shovkovy:2013, Preis:2011A,Preis:2011B, Sanfilippo:2010, catalysis:2014, Catalysis:2015}. The early, CME, is strongly related to the electric charge separation phenomenon which can be measured in HIC experiments such as ALICE at LHC \cite{ALICE:2012}, PHENIX \cite{PHENIX:2014} and STAR \cite{STAR:2009,STAR:2010,STAR:2011} at RHIC. The latter, the magnetic catalysis, is conjectured to largely influences our picture on the QCD phase-diagram. The way that the critical temperatures change with varying magnetic fields certainly remaps the boundaries separating hadrons and QGP and the freezeout parameters; the temperature and the baryon chemical potential.

Recently, different theoretical studies proposed various methods for the numerical calculations of the experimentally measured  magnetic field effects, such as lattice QCD simulations \cite{QCD:2012, lattice:2013b, QCD:2013c, QCD:2013d, lattice:2014}, hadron resonance gas (HRG) model \cite{HRG1,HRG2}, two-flavor Nambu-Jona-Lasinio model (NJL) \cite{Klevansky:1992, NJLsu2} and NJL with Ployakov loop fields \cite{Menezes:2009, Fukushima:2010l}, and PLSM  \cite{Mizher;2010, Skokov;2012, Ruggieri:2013, Tawfik:2014gga}. The QCD phase-diagram in external magnetic field \cite{Mizher;2010,Skokov;2012,QCD:2012} and squeezing QCD matter \cite{squeezing:2013} are examples on lattice simulations for QCD magnetic properties. Great details on understanding the phase structure of strongly interacting QCD matter in finite magnetic fields are reviewed in Refs. \cite{review1, review2,review3,Fraga:Chiral2008,Shovkovy:2013}. Moreover, other  models reveal interesting features about the response of finite magnetic field to hot and dense medium, such as higher-order moments of quark multiplicity \cite{Tawfik:2014uka}, chiral phase-structure of meson-states \cite{Tawfik:2014gga} and temperature dependence of some transport coefficients \cite{Tawfik:2016edq}.   Corrections to QCD-like models, such as LSM and NJL, should be checked for renormalization \cite{Caldas:2001A}. Dressing a scalar mass up to two-loop order at finite temperature was discussed in Ref. \cite{Caldas:2002}. 

To summarize, the present work utilizes the Polyakov linear-sigma model in order to analyze the magnetic properties of QCD matter in thermal medium. We present the temperature dependence of  magnetization, magnetic susceptibility,  permeability and  magnetic catalysis on finite magnetic field strength. Furthermore, we study the influence of finite magnetic field on the QCD phase-diagram and the interrelations between $(T\;\mbox{vs.}\;eB), \; (\mu\;\mbox{vs.}\;eB)$ and and $(T\;\mbox{vs.}\;\mu)$ QCD phase-diagrams. Following aspects belong to the main targets of this paper: 
\begin{enumerate}
\item characterizing the influences of finite magnetic field and Landau level quantization on the chiral quark-condensate and quark-hadron phase transitions, 
\item investigating the effects of finite magnetic field on the QCD phase-diagram, 
\item describing the magnetic catalysis in the QCD matter,  especially that we have obtained opposite results in a previous work \cite{Tawfik:2014gga}, which is compared with the present calculations, for instance middle panel of Fig. \ref{propes}.
\item whenever possible, confronting our calculations to recent first-principle lattice QCD simulations, and 
\item proposing possible signatures reflecting various magnetic properties of the QCD matter in thermal medium.
\end{enumerate}

The present paper is organized as follows. Short details about PLSM and its mean field approximation are introduced in section \ref{model}. An entire description on PLSM can be found in Refs. \cite{Tawfik:2016edq, Tawfik:2016ihn,Tawfik:2016lih, Tawfik:2014gga, Tawfik:2014uka}. In presence of finite magnetic field, great details about Landau quantization and the possible modifications on the PLSM partition-function shall be discussed in section \ref{sec:llquant}. The various order parameters of chiral quark-condensates and deconfinement order-parameters in a wide range of temperatures at different values of the magnetic field strengths are calculated in section \ref{Order:parameters}. In section \ref{magnetism}, some magnetic properties such as magnetization, magnetic susceptibility and permeability are compared with recent lattice QCD simulations. In section \ref{chiralMF}, we present the magnetic-field dependence of the critical temperature ($T_c$) and the baryon chemical potential ($\mu _c$) characterizing the chiral phase-transition. Also, we present the QCD phase-diagram in finite magnetic field. This section shall be followed by the conclusions in section \ref{conclusion}. 


\section{A short reminder to SU($3$) Polyakov linear-sigma model \label{model}}

The LSM Lagrangian with $N_f=3$  coupled to $N_c=3$ is given as $\mathcal{L} = \mathcal{L}_q+\mathcal{L}_m-\mathbf{\mathcal{U}}(\phi, \phi^*, T)$. 
\begin{itemize}
\item The first term defines the quark contributions, where quarks couple to mesons by flavor-blind Yukawa coupling $g$ \cite{blind,Caldas:2001A}, 
\begin{eqnarray}
\mathcal{L}_q &=& \sum_f \overline{q}_f(i\gamma^{\mu}
D_{\mu}-g\,T_a(\sigma_a + i \gamma_5 \pi_a))q_f, \label{lfermion} 
\eea
where $\mu,\; D_{\mu}$ and $\gamma^{\mu}$ are  an additional Lorentz index, covariant derivative and gamma matrices, respectively.
\item The second term stands to the meson contributions,
\begin{eqnarray}
\mathcal{L}_m &=&\mathrm{Tr}(\partial_{\mu} \Phi^{\dag}\partial^{\mu} \Phi-m^2
\Phi^{\dag} \Phi)-\lambda_1\,[\mathrm{Tr}\,(\Phi^{\dag} \Phi)]^2 \nonumber \\&&  
-\lambda_2\,\mathrm{Tr}(\Phi^{\dag} \Phi)^2+c[\mathrm{Det}(\Phi)+\mathrm{Det}(\Phi^{\dag})]
+\mathrm{Tr}[H(\Phi+\Phi^{\dag})],  \hspace*{6mm} \label{lmeson}
\end{eqnarray}
where $\Phi$ is $(3\times3)$ matrix includes the nonet meson states as
\bea
\Phi &=&  \sum_{a=0}^{N_f^2 - 1} T_a (\sigma_a -i \pi_a).
\eea 
The number of generators ($T_a$) is defined according to the number of quark flavors ($N_f$). In  U$(3)$ algebra, $T_a$ is determined by Gell-Mann matrices $\hat{\lambda}_a$ \cite{Weinberg1972}; $T_a= \hat{\lambda}_a/2$ with $a=0,\cdots,\,8$.  Tab. \ref{tab:1a} summarizes the values of the parameters,  $m^2$, $h_l$, $h_s$, $\lambda_1$, $\lambda_2$, and $c$. These six values are estimated at sigma mass $m_\sigma=800~$MeV \cite{Schaefer:2008hk}. 

\begin{table}[htb]
\begin{center}
\begin{tabular}{|c | c | c | c | c | c | c |}
\hline
$m_\sigma$ [MeV] & $c\,$ [MeV] & $h_l\,$ [MeV$^3$] & $h_s\,$ [MeV$^3$] & $m^2 \,$ [MeV$^2$] & $\lambda _1$ & $\lambda _2$\\ 
\hline 
800 & $4807.84$ & $(120.73)^3$ & $(336.41)^3$ & -$(306.26)^2$ & $13.49$& $46.48$\\ 
\hline 
\end{tabular}
\caption{Summary of PLSM's parameters.  A detailed description is given in Ref. \cite{Schaefer:2008hk}  \label{tab:1a}}
\end{center}
\end{table} 

%
\item The third term, the potential $\mathcal{U}(\phi, \phi^{*},T)$, gives  Polyakov-loop potential, which introduces the dynamics of gluons and the quark interactions. The Polyakov-loop variables are motivated by the underlying QCD symmetries in the pure gauge theory  \cite{Caldas:2001A}. This potential can be adjusted from recent lattice QCD simulations and likely has $Z(3)$ center-symmetry \cite{Ratti:2005jh, Roessner:2007, Schaefer:2007d, Fukushima:2008wg}. Through the thermal expectation value of a color-traced Wilson-loop in the temporal direction, the dynamics of color charges and gluons are taken into consideration 
\begin{eqnarray}
\phi = \langle\mathrm{Tr}_c\, \mathcal{P}\rangle/N_c, \qquad && \qquad
\phi^* = \langle\mathrm{Tr}_c\,  \mathcal{P}^{\dag}\rangle/N_c. \label{phis}
\end{eqnarray}

There are various proposals for the Polyakov-loop potentials. In the present work, we utilize the polynomial form for Polyakov variables ($\phi$ and $\phi^{*}$)  \cite{Ratti:2005jh, Roessner:2007, Schaefer:2007d, Fukushima:2008wg}, 
\begin{eqnarray}
\mathbf{\mathcal{U}}(\phi, \phi^*, T)= T^4 \left\{-\frac{b_2(T)}{2}|\phi|^2-\frac{b_3}{6}\left(\phi^3+\phi^{*3}\right)+\frac{b_4}{4}\left(\left| \phi \right|^2\right)^2 \right\}, \label{Uloop}
\end{eqnarray}
where $b_2(T)=a_0+a_1\left(T_0/T\right)+a_2\left(T_0/T\right)^2+a_3\left(T_0/T\right)^3$. For a good agreement with the lattice QCD results, the deconfinement temperature of pure gauge $T_0=270$ MeV and  $a_0=6. 75$,  $a_1=-1. 95$,  $a_2=2. 625$, 
 $a_3=-7. 44$, $b_3=0.75$, $b_4=7.5 $  are used  \cite{Ratti:2005jh}. 
\end{itemize}

In thermal equilibrium, the grand-canonical partition function ($\mathcal{Z}$) at finite $T$ and $\mu_f$, where the subscript $f$ refers to quark flavors, can be constructed.  At finite volume ($V$), the free energy is given as  $\mathcal{F}=-T \cdot \log [\,\mathcal{Z}]/V$  or 
\begin{equation}
\mathcal{F}  
= U(\sigma_l, \sigma_s) +\mathbf{\mathcal{U}}(\phi, \phi^*, T) + \Omega_{  \bar{q}q}(T, \mu _f, B)  + \delta_{0,eB} \,\Omega_{ \bar{q}q}(T, \mu _f), \label{potential}
\end{equation}
where the last two terms represent the quark-antiquark contributions at finite and vanishing magnetic field, respectively. $ \delta_{0,eB}$ switches between both terms. Practically, only one of them shall be taken into account, separately.
\begin{itemize}
\item Assuming that the nonstrange (light) and strange quark condensates are given as $\sigma_l$ and $\sigma_s$, respectively, the purely mesonic potential reads 
 \begin{eqnarray}
U(\sigma_l, \sigma_s) &=& - h_l \sigma_l - h_s \sigma_s + \frac{m^2}{2}\, (\sigma^2_l+\sigma^2_s) - \frac{c}{2\sqrt{2}} \sigma^2_l \sigma_s  \nonumber \\&& 
+ \frac{\lambda_1}{2} \, \sigma^2_l \sigma^2_s +\frac{(2 \lambda_1 +\lambda_2)}{8} \sigma^4_l  + \frac{(\lambda_1+\lambda_2)}{4}\sigma^4_s. \hspace*{8mm} \label{Upotio}
\label{pure:meson}
\end{eqnarray}

\item In nonzero magnetic field  ($eB \neq 0$) and at finite $T$ and $\mu_f$,  the concepts of Landau quantization and magnetic catalysis, where the magnetic field is assumed to be oriented along $z$-direction, should be implemented, properly.  The Landau-level structure is conjectured to have effect on the phase space \cite{Shovkovy:2013}, 
\bea
 \int \frac{d^3p}{(2\pi)^3}  \longrightarrow   \frac{|q_f|B}{2\pi} \sum_\nu \int \frac{dp_z}{2\pi} (2-\delta_{0\nu}), \label{phaseeB}
\eea
where $\nu$ gives the Landau quantization levels, section \ref{sec:llquant}.  The quark and antiquark contributions to the potential are given as
\begin{eqnarray} 
\Omega_{ \bar{q}q}(T, \mu _f, B)&=& - 2 \sum_{f=l, s} \frac{|q_f| B \, T}{(2 \pi)^2} \,  \sum_{\nu = 0}^{\nu _{max_{f}}}  (2-\delta _{0 \nu })    \int_0^{\infty} dp_z \nonumber \\ && \hspace*{5mm} 
\left\{ \ln \left[ 1+3\left(\phi+\phi^* e^{-\frac{E_{B, f} -\mu _f}{T}}\right)\; e^{-\frac{E_{B, f} -\mu _f}{T}} +e^{-3 \frac{E_{B, f} -\mu _f}{T}}\right] \right. \nonumber \\ 
&& \hspace*{3.7mm} \left.+\ln \left[ 1+3\left(\phi^*+\phi e^{-\frac{E_{B, f} +\mu _f}{T}}\right)\; e^{-\frac{E_{B, f} +\mu _f}{T}}+e^{-3 \frac{E_{B, f} +\mu _f}{T}}\right] \right\}, \label{PloykovPLSM}
\end{eqnarray}
where $E_{B,f}$ is the dispersion relation of $f$-th quark-flavor in finite magnetic field, Eq. (\ref{eq:moddisp}).  Other modifications shall be discussed in section \ref{sec:llquant}.
\item At zero magnetic field ($eB=0$) and finite temperature ($T$) and chemical potential ($\mu_f$)
\begin{eqnarray} 
\Omega_{ \bar{q}q}(T, \mu _f)&=& -2 \,T \sum_{f=l, s} \int_0^{\infty} \frac{d^3\vec{P}}{(2 \pi)^3} \left\{ \ln \left[ 1+3\left(\phi+\phi^* e^{-\frac{E_f-\mu _f}{T}}\right)\times e^{-\frac{E_f-\mu _f}{T}}+e^{-3 \frac{E_f-\mu _f}{T}}\right] \right. \nonumber \\ 
&& \hspace*{35.5mm} \left.  +\ln \left[ 1+3\left(\phi^*+\phi e^{-\frac{E_f+\mu _f}{T}}\right)\times e^{-\frac{E_f+\mu _f}{T}}+e^{-3 \frac{E_f+\mu _f}{T}}\right] \right\}, \hspace*{8mm} \label{PloykovPLSM}
\end{eqnarray}
where $E=\sqrt{\vec{P}^2+m_f^2}$ is the dispersion relation of valence quark and antiquark and $m_f$ is the $f$-th mass of quark flavour.
\end{itemize}

As discussed in Ref. \cite{Tawfik:2014uka}, the temperature dependence of the LSM mesonic potential becomes considerable at low temperatures. At higher temperatures, this weakens, exponentially. Accordingly, the corresponding term in Eq. (\ref{potential}) can be excluded as given in Refs. \cite{Tawfik:2014gga, Tawfik:2014uka}. The presence of the LSM mesonic potential  is necessary in order to introduce the chiral symmetry breaking and the mesonic fluctuations.    
%
The Polyakov loops can be added to the model through gluonic potential, $\mathbf{\mathcal{U}}(\phi, \phi^*, T)$. By doing that, the dynamics of gluons shall be taken into consideration. When confronting our calculations to recent lattice simulations, section \ref{resulat}, we observe that both are in a good agreement. This would be explained because our PLSM is so configured that it integrates degrees of freedom, symmetries, and dynamics, etc. that enable PLSM to fit well with the first-principle lattice simulations. Nevertheless, PLSM remains an effective model to QCD.

In order to evaluate the expectation values of chiral quark-condensates, $\sigma_l$ and $\sigma_s$, and deconfinement order parameters $\phi$ and $\phi^*$, one can minimize the free energy $\mathcal{F}$ at finite volume, Eq. (\ref{potential}),
\begin{eqnarray}
\left.\frac{\partial \mathcal{F} }{\partial \sigma_l}\right|_{min}, 
\left.\frac{\partial \mathcal{F}}{\partial \sigma_s}\right|_{min},
\left.\frac{\partial \mathcal{F}}{\partial \phi}\right|_{min},
\left.\frac{\partial \mathcal{F} }{\partial \phi^*}\right|_{min}. \label{cond1}
\end{eqnarray}
In finite chemical potential $\mu \ne 0$, the PLSM free energy at finite volume, Eq. (\ref{potential}), becomes complex. Therefore, the analysis of PLSM order-parameters is given by minimizing the real part of free energy, i.e. Re $\mathcal{F}$. Solutions for PLSM order-parameters can be evaluated by minimizing the real part of $\mathcal{F}$ at the saddle point. At finite magnetic field, the temperature and dense dependences can be estimated. Concretely, the order parameters $\sigma_l=\bar{\sigma_l}$, $\sigma_s=\bar{\sigma_s}$, $\phi=\bar{\phi}$ and $\phi^*=\bar{\phi^*}$ and their dependences on $T$, $\mu$ and $e B$ can be evaluated. To assure minimal $\mathcal{F}$, one can illustrate this graphically and/or evaluate its second derivative. We have conducted both (not shown here).

It is worthwhile highlighting that the adjustment of the pure gauge potential to Polyakov-loop potential improves the chiral model towards best agreement with recent lattice QCD simulations. Nevertheless, the construction of PLSM allows to describe the quark-hadron phase structure, where the valence and sea quarks are implemented, Eq. (\ref{potential}). The mechanism of the magnetic catalysis relies on a competition between the contributions of valance and sea quarks \cite{Bruckmann:2013, Fraga:PLB2014}. In light of this, the influence of finite magnetic field implies a suppression on the quark condensates (sea quarks) leading to a net inverse magnetic catalysis. Furthermore, the valence quark potential has a very small effect on the free energy, especially at high temperature, Eq. (\ref{potential}). This makes the contributions of the sea quarks more dominant than that of the valence quarks. The contributions of sea quarks can be considered as a backreaction of the quarks in pure gauge fields \cite{Anders:JHEP2004}. If this backreaction is incorporated in the model, one shall be able to find good agreements with lattice QCD simulations. Nevertheless, the agreement reported in the present paper, apparently means that the PLSM assumes - among others - correct degrees of freedom in both hadronic and partonic phases. 

For the sake of completeness, we highlight that removing the ultraviolet divergences can be achieved through the fermion vacuum term (the zero Matsubara mode) \cite{Redlich:2010d}. Nevertheless, we found (not shown here) that  its negligible value, especially at the temperatures defining the scope of this work \cite{Menezes:2009a}. For sharp noncovariant cut-off $(\Lambda)$
\bea
\Omega^{\mbox{vac}}_{q\bar{q}} &=&  2 N_c N_f \sum_f \int_\Lambda  \frac{d^3P}{(2\pi)^3}  E_f = \frac{-N_c N_f}{8\pi^2} \sum_f \left( m_f^4 \ln {\left[\frac{\Lambda + \epsilon_\Lambda}{m_f} \right]} - \epsilon_\Lambda \left[\Lambda^2 + \epsilon_\Lambda^2\right]\right),
\eea
where $\epsilon_\Lambda=(\Lambda^2 + m_f^2)^{1/2}$ and $m_f$ is the mass of $f$-th quark flavor.

\section{Landau quantization \label{sec:llquant}}
 As discussed, Landau quantization is an essential consequence of applying finite magnetic field to PLSM. Accordingly, we have to analysis how the different Landau levels are populated. Such a consequence appears in the dispersion relation, which should be modified at finite magnetic field,
\bea
E_{B, f}(B)&=&\left[p_{z}^{2}+m_{f}^{2}+|q_{f}|(2n+1-\sigma) B\right]^{1/2}, \label{eq:moddisp}
\eea 
with $n$, a quantization number, is known as the Landau quantum number and $\sigma$ is related to the spin quantum number, $\sigma=\pm S/2$ and to the masses of quark-flavor, where $f=l,s$ with $l$ runs over $u$ and $d$ quarks and the other subscript ($s$) stands for $s$-quarks.  Furthermore, Landau quantization enters the summation in Eq. (\ref{PloykovPLSM}). Accordingly, we highlight the crucial importance of the so-called zero-level, which considerably differs from $\Omega_{\bar{q}q}$ at $e B=0$ and greatly responsible whether or not the QCD matter possesses direct or indirect magnetic catalysis. 

The quark masses are directly coupled to the corresponding sigma fields
\bea
m_l = g\, \frac{\sigma_l}{2}, \qquad & & \qquad
m_s = g\, \frac{\sigma_s}{\sqrt{2}}.  \label{qmassSigma}
\eea
The quantity $2n+1-\sigma$ can be replaced by a sum over the Landau Levels; $0\, \leq \nu \, \leq \nu_{max_f}$. The lower bound, in this inequality, is the Lowest Landau Level, while the higher one stands for the Maximum Landau Level ($\nu_{max}$). For the sake of completeness, we mention that $2-\delta_{0 \nu}$ represents degenerate Landau Levels. $\nu_{max_{f}}$ contributes to the maximum quantization number ($\nu_{max_{f}} \rightarrow \infty$). 

Thus, a considerable influence of  the baryon chemical potential, the temperature and the magnetic fields on the number of Landau levels should be taken into account. One of such proposal was introduced in Ref. \cite{Boomsma:2010}, 
\begin{equation}
\nu_{max_{f}} = \left\lfloor  \frac{\tau_f ^2 - \Lambda ^2 _{QCD}}{2 |q_f| B } \right\rfloor, \label{MLL}
\end{equation} 
where the brackets represent floor of the enclosed quantity. The parameter $\tau_f$ is conjectured to be related to the baryon chemical potentials of $f$-th quark flavor \cite{Boomsma:2010}. To avoid confusion, this expression can be omitted, especially that it was now utilized in our calculations. In next sections, we shall elaborate a short summary of our results on MLL at varying $T$, $\mu$, and $eB$.

A systematic study for the Landau levels occupied by the quarks is now in order. This differs from a quark flavor to another and apparently varies with the magnetic field and the baryon chemical potential. According to Eq. (\ref{MLL}), the maximum occupation number of the Landau level depends on the quark charges, the magnetic fields, the temperatures, and the baryon chemical potentials. Maximum Landau levels (MLL) for quarks should have different occupations according to the large change in the quark charges.  For example, at $eB=m^2_{\pi}$, the maximum Landau levels (MLL) depend on the baryon chemical potential as follows.
\begin{itemize}
\item At $\mu=0~$ and $100~$MeV, the up-quarks have up to $62$ levels, while the down- and the strange-quarks occupy $124$ levels each.
\item At $\mu=200~$MeV, the up-quarks can fill up $59$ levels, while the down- and the strange-quarks each is allowed to accommodate $118$ Landau levels.
\end{itemize} 
In such way, the MLL at $eB=10\, m^2 _{\pi}$ can be counted as,
\begin{itemize}
\item At $\mu=0$ and $100~$MeV and for up-quarks, MLL$=3$, while and for each down- and strange-quarks MLL$=6$.
\item At $\mu=200~$MeV,  for up-quarks MLL becomes $2$, while for each down- and strange-quarks, MLL can be as much as $4$. 
\end{itemize}

We conclude that increasing the magnetic field leads to filling-up the lower Landau levels first and decreasing the number of occupied levels. In other words, increasing the magnetic field allows lower Landau level to accommodate more quarks.

Furthermore, in Eq. (\ref{MLL}), one can replace the chemical potential ($\mu_f$) by the temperature ($T$). We found that the population of MLL depends on the temperature, the quark charge, and the magnetic field strength. As given in Eq. (\ref{MLL}), this can be  scaled by $\Lambda_{QCD}$. The main difference between MLL occupation of up- and down-quark is that $|q_d|= 2 |q_u|$. 
\begin{itemize}
\item At $T=50\,$MeV and $eB=m^2_{\pi}$, the up-quark has $31$ MLL, while each down- and strange-quark has $62$ MLL.
\item At $T=100\,$MeV and $eB=15\, m^2_{\pi}$, MLL for up-quark  is $2$ and $4$ for each of down- and strange-quark. 
\end{itemize}
To summarize, we can conclude that the population of the Landau levels is most sensitive to the magnetic field and to the quark charges. Also, MLL is strongly controlled by the QCD scale ($\Lambda_{QCD}$). We assure that our calculations assume maximum population of the Landau levels, except the order parameters in Fig. \ref{fig:sbtrc3}. They are estimated at varying occupations of the Landau levels.

\section{The results} 
\label{resulat}

The chiral quark-condensates and deconfinement order-parameters shall be analyzed in a wide range of temperatures, baryon chemical potentials, magnetic fields, and at different populations of the quantized Landau levels. The temperature and density dependence of some magnetic properties such as magnetization, magnetic susceptibility, and permeability shall be determined in finite magnetic fields. Furthermore, the {\it magnetic} phase-diagram, i.e.  the variation of the critical temperatures ($T_c$) and the corresponding (critical) baryon chemical potential ($\mu _c$) and finite magnetic fields shall be studied. Concretely, the QCD phase-diagram ($T$ vs. $\mu$) at different magnetic field strengths shall be mapped out. The variation of temperature, baryon chemical potential and magnetic field from {\it ordinary} chemical freezeout conditions such as constant normalized entropy density, $s/T^3=7$, shall be presented. 

First, we introduce the chiral quark-condensates, the deconfinement order-parameters and magnetic catalysis in thermal medium in the section that follows.

\subsubsection{Chiral quark-condensates, deconfinement order-parameters and magnetic catalysis}
\label{Order:parameters}

The chiral quark-condensates ($\sigma_l$ and $\sigma_s$) and the deconfinement order-parameters ($\phi$ and $\phi^*$) in dense and thermal medium are estimated through the so-called {\it global} minimization of the free energy, Eq. (\ref{cond1}). In the present work, the parameters of PLSM are estimated at sigma-meson mass $m_{\sigma}=800~$MeV, the vacuum mass, where the measured (vacuum) light and strange chiral condensates are assumed as $\sigma_{l_o}=92.5~$ MeV and $\sigma_{s_o}=94.2~$MeV, respectively. These parameters are partly responsible for the excellent agreement with the first-principle lattice simulations, about which we shall report in forthcoming sections.

\begin{figure}[htb]
\centering{
\includegraphics[width=5.cm,angle=-90]{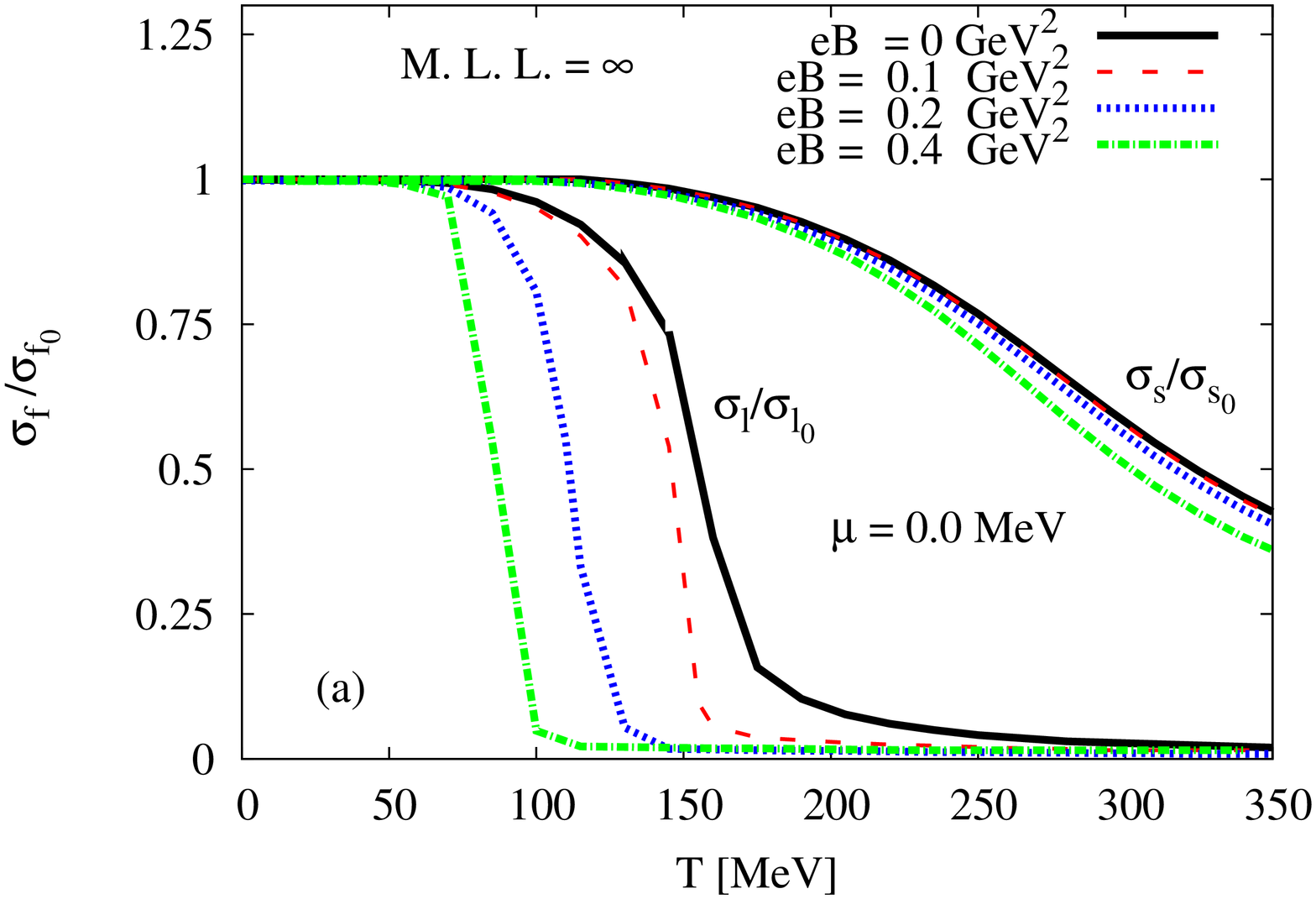}
\includegraphics[width=5.cm,angle=-90]{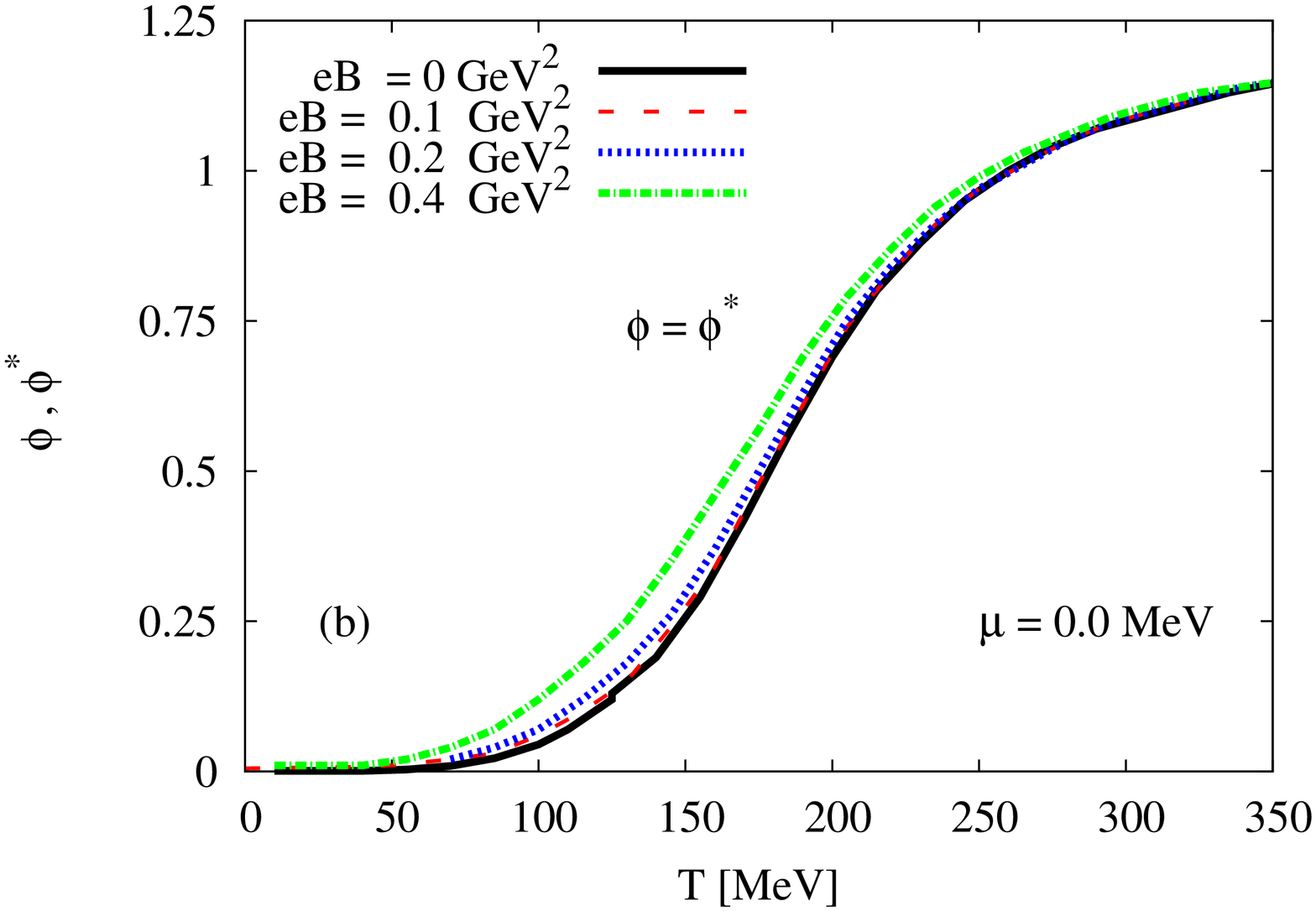}
\caption{(Color online)  Left-hand panel: the chiral quark-condensates normalized to the corresponding vacuum values are given as functions of temperature at vanishing baryon chemical potential  and $eB=0$ (solid curves), $0.1$ (dashed curves), $0.2$ (dotted curves) and $0.4~$GeV$^2$ (dot-dashed curves) and at vanishing baryon chemical potential. Right-hand panel: the same as in the left-hand panel but for  expectation values of Polyakov-loop fields ($\phi$  and $\phi^*$). \label{fig:sbtrc2}
}}
\end{figure} 

Figure \ref{fig:sbtrc2} shows the temperature dependence of normalized chiral  quark-condensates (a) and deconfinement order-parameters (b) at different magnetic field strengths; $eB= 0$ (solid curves), $0.1$ (dashed curves), $0.2$ (dotted curves) and $0.4~$GeV$^2$ (dot-dashed curves) and at vanishing baryon chemical potential. In left-hand panel (a), we notice that the chiral critical temperature decreases with increasing the magnetic field. This means that the phase transition known as crossover becomes sharper with increasing the magnetic field. This can be interpreted due to the maximum occupation of the Landau levels ($\nu_{max}\rightarrow \infty$). So far, we conclude that the phase transition seems to be of first order whenever the chiral condensate passes through a metastable phase, in which light quarks become massless and move freely.

In right-hand panel (b), the temperature dependence of the deconfinement order-parameters is depicted at a vanishing baryon chemical potential, i.e. $\phi=\phi^*$ but different values magnetic field strengths; $eB=0$ (solid curves), $0.1$ (dashed curves), $0.2$ (dotted curves) and $0.4~$GeV$^2$ (dot-dashed curves). It is obvious that the deconfinement critical-temperature ($T_{\phi}$) very slightly decreases as the magnetic field increases.

\begin{figure}[htb]
\centering{
\includegraphics[width=5.cm,angle=-90]{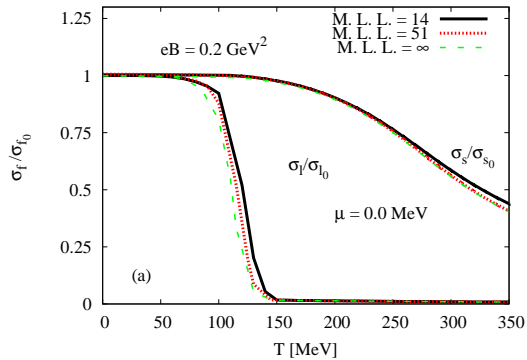}
\caption{(Color online) The normalized chiral quark-condensates are given as functions of temperature at finite magnetic field $eB=0.2~$GeV$^2$ but different populations for the maximum Landau levels. \label{fig:sbtrc3}
}}
\end{figure}

As discussed in section \ref{model}, PLSM is well suited to study the chiral limit. The inclusion of magnetic field in PLSM can be partly achieved by changing the dispersion relation for quarks and antiquarks,  Eq. (\ref{eq:moddisp}). In doing this, the dimension of the momentum-space should be reduced (from three to one) and scaled via quark charge and magnetic field. This process is known as dimension reduction or magnetic catalysis effect \cite{Shovkovy:2013}. Furthermore, the introduction of the magnetic field requires suitable implementation of the Landau quantization. 

Figure \ref{fig:sbtrc3} shows - for the first time - the effects of the occupation of the Landau levels on the temperature dependence of the  chiral quark-condensates ($\sigma_l$ and $\sigma_s$) at a finite magnetic field $eB=0.2~$GeV and a vanishing baryon chemical potential. We observe that the change in the Landau levels is only relatively significant within the phase transition and seems to disappear otherwise. At $MLL=14$ (solid curves), $51$ (dotted curves) and $\infty$ (dashed curves), the normalized chiral condensates for light and strange quarks are analyzed as functions of temperature at a finite magnetic fields and a vanishing baryon chemical potential. We conclude that increasing the Landau levels very slightly sharpens the phase transition and decreases the critical temperature $T_{\chi}$. The latter characterizes an inverse magnetic catalysis.

Some fundamental properties of strongly interacting QCD matter in thermal medium and at finite magnetic field such as magnetization, magnetic susceptibility and permeability shall be estimated in the following section.

\subsubsection{QCD magnetization, magnetic susceptibility and permeability}
\label{magnetism}

The magnetic susceptibility with proper renormalization has been introduced in Ref. \cite{susceptibility:2014}. The quantity estimates the ability of the QCD matter to generate the magnetic field. In another words, it measures the ability to store magnetic potential energy, which is defined as a proportionally constant for the magnetic flux. The magnetic flux is formed or produced from the influence of the magnetic field. The magnetic permeability is calculated along the magnetic field that aligns on the transverse direction to the momentum space $p_z$. The strong magnetic field likely results in isotropic QCD matter. 

The response of  the QCD matter to an external magnetic field can be estimated from the free energy density $\mathcal{F}= - T/V \cdot \ln \mathcal{Z}$. In thermal, dense and magnetic medium, the partition function $\ln\, \mathcal{Z}$ gets modifications, from which the magnetization can be deduced 
\begin{equation}
\mathcal{M}=- \frac{\partial \mathcal{F}}{\partial (eB)}, 
\end{equation}
where $e\neq 0$ is the elementary electric charge. In natural units, the magnetization is given in GeV$^2$. The sign of magnetization determines an important magnetic property; whether QCD matter is {\it para}- or {\it dia}-magnetic, i.e.  $M>0$ (para-), or $M<0$ (dia-), respectively. As in solid-state physics, 
\begin{itemize}
\item if the QCD matter is in state of {\it dia}-magnetization, the color charges align oppositely to the direction of the magnetic field and produce an induced current, which spreads as small loops attempting to cancel out the effects of the applied magnetic field, and 
\item if the QCD matter is in state of  {\it para}-magnetization, the most color charges align towards the direction of the magnetic field. 
\end{itemize}

 Let us first recall the classical electromagnetism! It is known that the magnetization diminishes with increasing temperature. Accordingly, the magnetic susceptibility ($\chi=c/T$, where $c$ is the Curie's constant) depends on the magnetic permeability ($\mu_B$) \cite{Hall}. This means that - in classical theory - the temperature has an inverse effect on the magnetization. Thus,  as per classical theory, the magnetization vanishes at very high temperature. This temperature limit in the strong interactions is likely at relativistic energies or at vanishing baryon chemical potential.  In this regard, we have to distinguish between the various magnetic properties of the strongly interacting QCD matter in thermal medium not only by determining the magnetization. What we observed points out to an opposite temperature-dependence of QCD-magnetization. The magnetic susceptibility and permeability play an essential role. In other words, the response of the QCD matter to finite magnetic field can be determined by the slope of magnetization ($\mathcal{M}$) with respect to the magnetic field. 

The second derivative of free energy density with respect to finite magnetic field results in the magnetic susceptibility
\begin{equation}
\chi_B= - \frac{\partial^2 \mathcal{F}}{\partial (eB)^2}\biggr\rvert_{eB=0}.
\end{equation} 
The magnetic susceptibility is a dimensionless proportionality parameter indicating the degree of magnetization of the QCD matter. 

Furthermore, the relative magnetic permeability ($\mu _{r}$) relative to the vacuum one $\mu_0$ can be translated as the magnetic effect in thermal QCD medium. This can be  determined by different methods such as direct relation with the magnetic susceptibility 
\begin{equation}
\mu_{r} = 1 + \chi_B.
\end{equation} 
This general formula is very common in solid-state materials. As shall be introduced in the following sections, this relation agrees well with the lattice QCD simulations, in which the magnetic permeability is expressed in terms of the magnetic susceptibility
 \begin{equation}
 \mu_B \equiv \frac{B^{ind}}{B^{ext}}  = \frac{1}{1 - 4 \pi \alpha _{m} \cdot \chi _B}, \label{permeability}
 \end{equation}
where $\alpha _{m}=e^2 /4\pi$ is the fine structure constant. This expression distinguishes between external $B^{ext}$ and the induced magnetic field $B^{ind}$. Both quantities are dimensionless proportionality constants. One remark on Eq. (\ref{permeability}) is now in order. The higher-order permeability seems to be limited by the magnetic susceptibility, which is given by the reciprocal of the square of elementary charge $e$, i.e.   $\chi_B \xrightarrow{\scriptscriptstyle \mu\to\infty} 1/e^2$.

\begin{figure}[htb]
\centering{
\includegraphics[width=3.9cm,angle=-90]{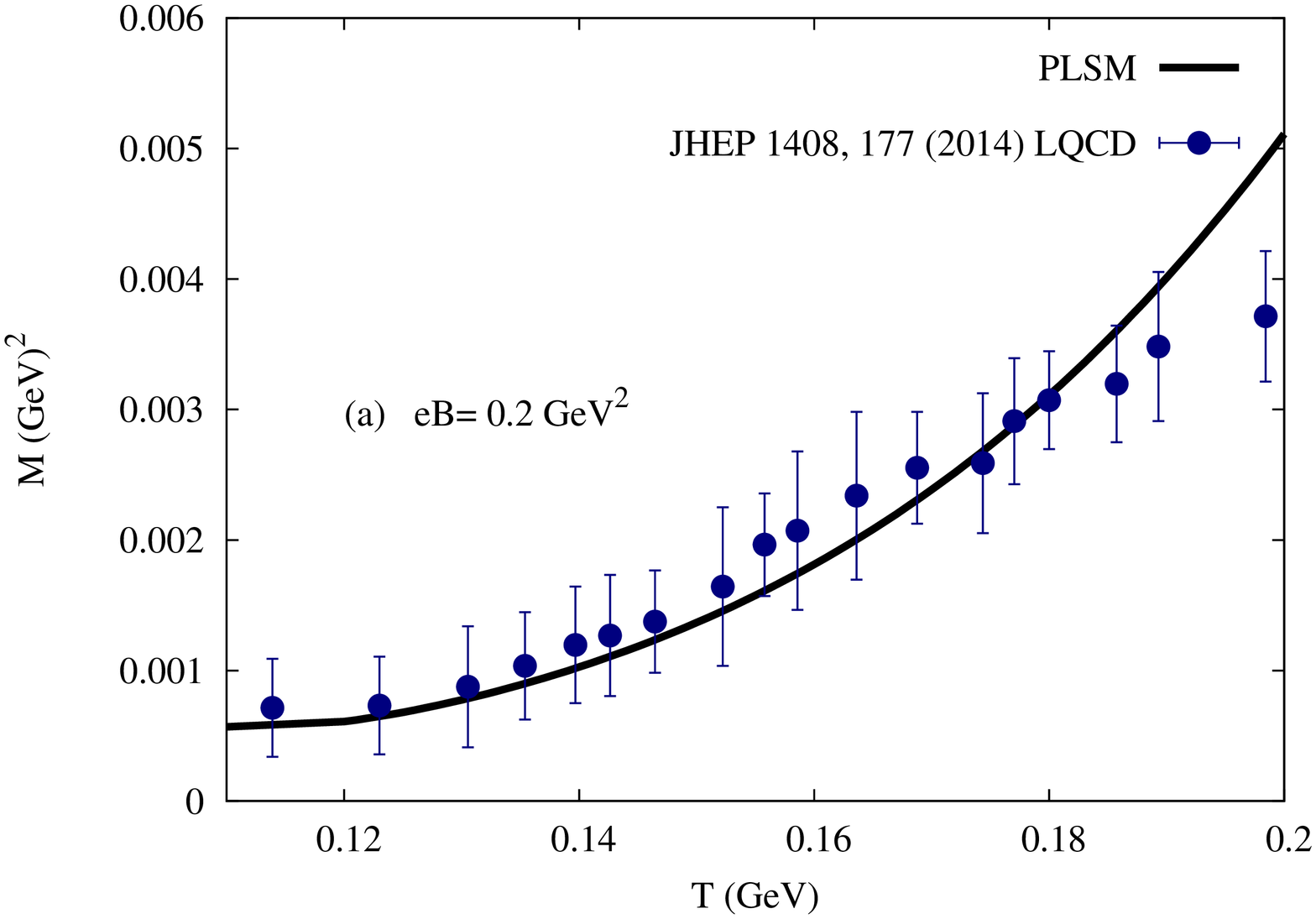}
\includegraphics[width=3.9cm,angle=-90]{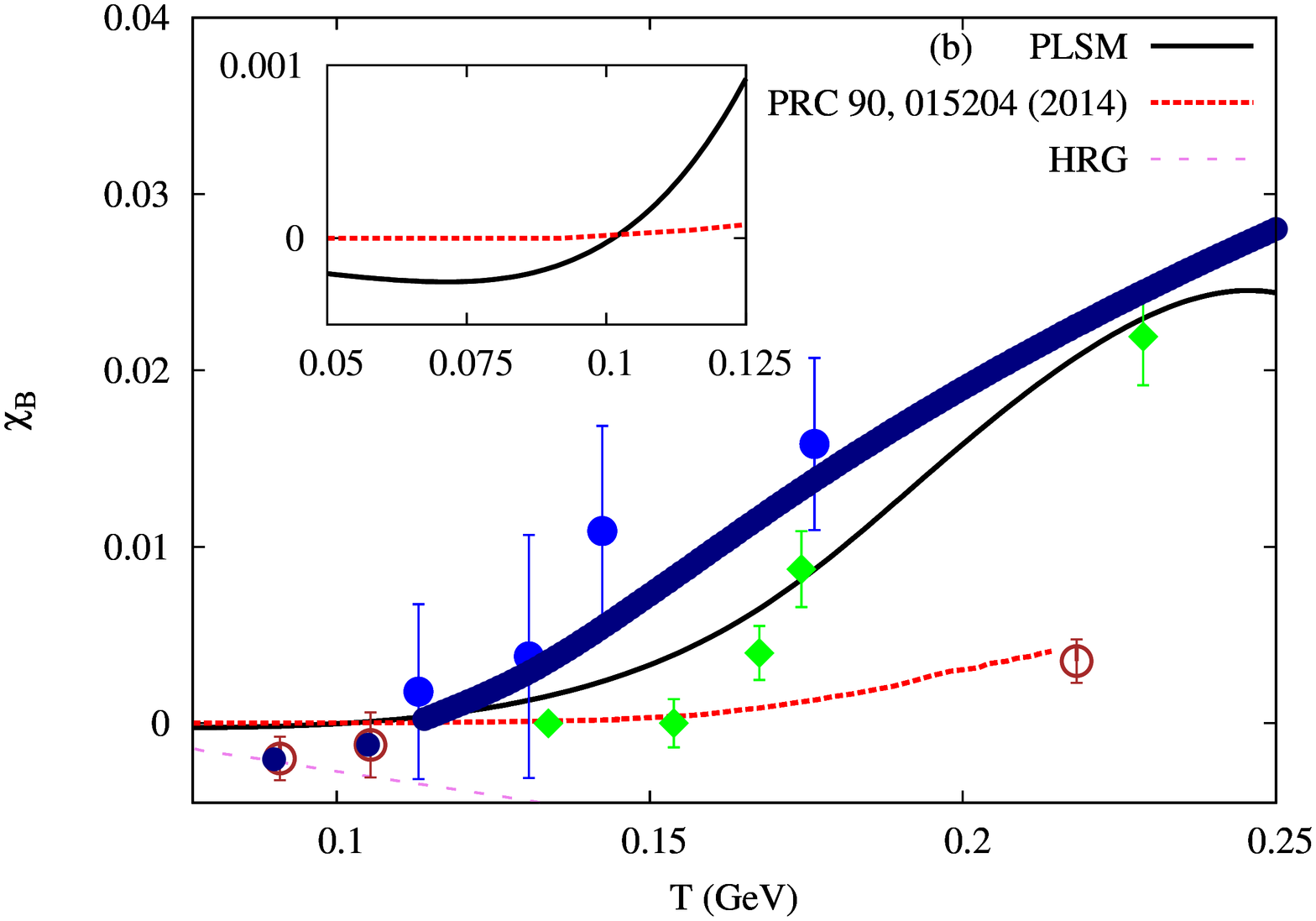}
\includegraphics[width=3.9cm,angle=-90]{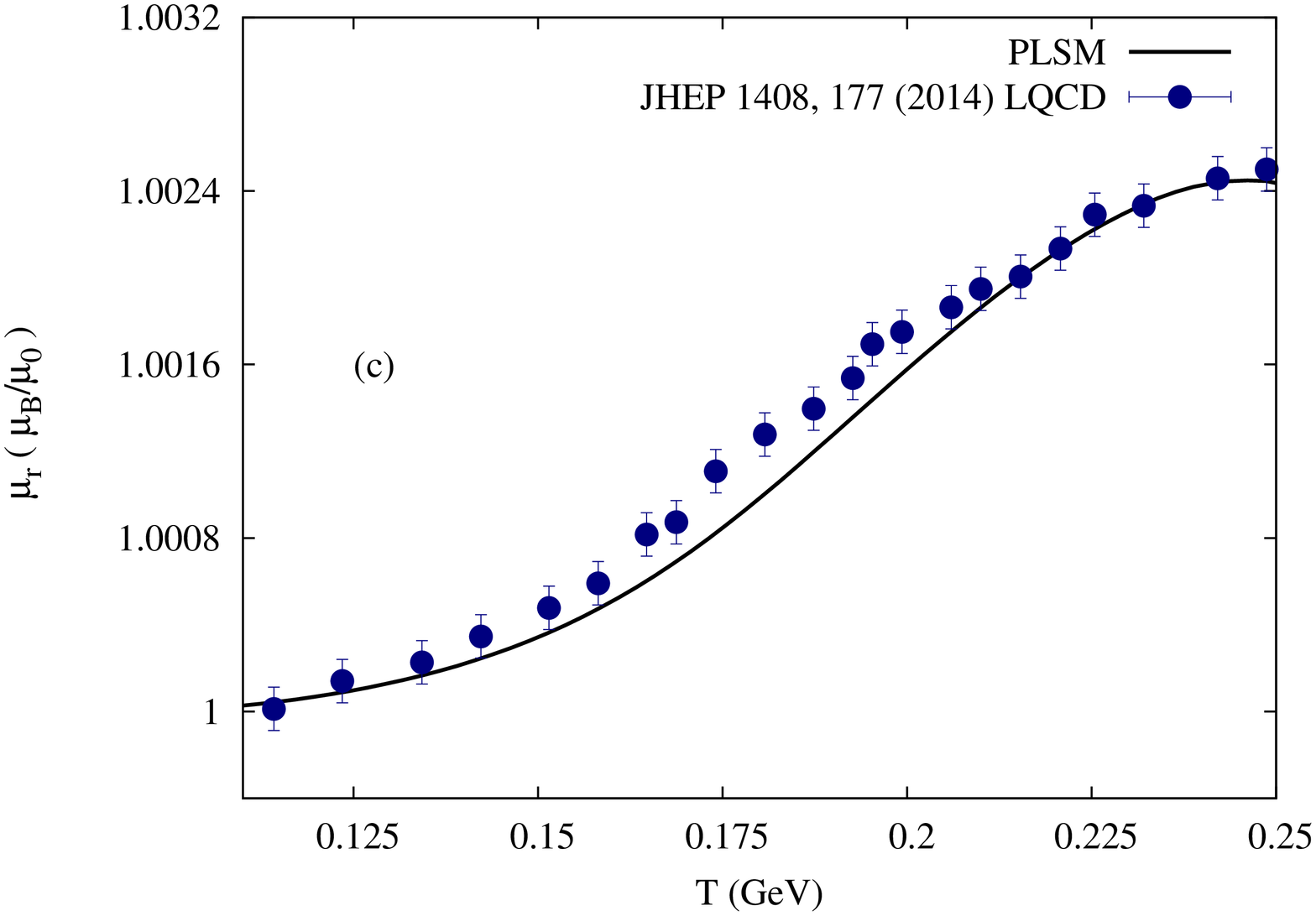}
\caption{(Color online) At $eB=0.2~$GeV$^2$ and vanishing baryon chemical potential, the temperature dependence of magnetization $\mathcal{M}$ (left-hand panel), magnetic susceptibility $\chi$ (middle-panel) and relative magnetic permeability (right-hand panel) is depicted. The results are compared with different lattice simulations (symbols) \cite{lattice:2014}.  The dashed curve represents our old calculations as reported in Ref. \cite{Tawfik:2014gga}.  \label{propes}
}}
\end{figure}

As mentioned earlier, the calculations from PLSM are in a good agreement with recent lattice QCD simulations. It is believed that such a comparison might lead to developing an intuitive understanding about the QCD matter in magnetic and thermal medium. In Fig. \ref{propes}, the magnetic properties of the QCD matter such as magnetization (left-hand panel), magnetic susceptibility (middle panel) and permeability (right-hand panel) are given as functions of temperature at nonvanishing magnetic field strength but a vanishing baryon chemical potential. The PLSM results (curves) are compared with various lattice QCD calculations (symbols). There is a good agreement over a wide range of temperatures.

In left-hand panel of Fig. \ref{propes}, the magnetization of the QCD matter in units of GeV$^2$ due to effects of nonvanishing magnetic field $eB =0.2~$GeV$^2$ and $\mu=0.0$ is studied as a function of temperature and compared with recent lattice calculations (open triangles with errorbars) \cite{lattice:2014}. The positive slope (or increasing magnetization with increasing temperature) refers to positive magnetization, $M>0$, which  indicates that the paramagnetic contribution of the QCD matter becomes dominant. Within the temperature range characterizing the hadron phase (below critical temperature), the curve seems to resemble the lattice calculations in an excellent way. At temperatures characterizing QGP (above critical temperature), the PLSM curve becomes larger than the lattice results, especially at very high temperatures. In this range of temperatures, the hadrons are conjectured to deconfine into color charges, quarks and gluons degrees of freedom.  It is apparent that such degrees of freedom are not sufficient enough to achieve a good agreement at very high temperature. Furthermore, the applicability of PLSM, which is mainly determined by the dominance (validity) of $\sigma_l$, $\sigma_s$, $\phi$ and $\phi^*$ order parameters at temperature, baryon chemical potential and magnetic field, section \ref{Order:parameters}, seems to reach an end at very high temperature. Some details about the lattice QCD simulations \cite{lattice:2014} are now in order. The  results on the magnetization at $eB \approx 0.2~$GeV$^2$ are obtained by using  half-half method for three lattice spacings employed at $N_\tau =6,$ and $8$ and continuum estimates. This is a partial explanation for the given results.

The middle-panel (b) of Fig. \ref{propes} shows the magnetic susceptibility as a function of temperature. The results from PLSM are compared with various lattice simulations (symbols) using different calculation methods and with the HRG calculations. The dashed curve stands for the old calculations  reported in Ref. \cite{Tawfik:2014gga}. These are greatly distinguishable from the new calculations. The reason is the exclusion of zero Landau levels and the mesonic contributions, when performing the old calculations \cite{Tawfik:2014gga}, the absence of this term leads to ignoring the mesonic fluctuations \cite{Kamikado2015}. In the present work, we include such fluctuations, Eqs. (\ref{potential}) and (\ref{PloykovPLSM}). Furthermore, The implementation of both conditions, i.e. the lowest Landau level and the mesonic potential) - in the present calculations -  makes the model agreeing well with the recent lattice QCD simulations, especially regarding the inverse magnetic catalysis and the weak evidence of dia-magnetic property of thermal QCD matter.

There is qualitative and quantitative agreement between our PLSM calculations and lattice QCD. The lattice QCD calculations among themselves have large differences. Thus, some features on PLSM and lattice QCD results can be summarized as follows.
\begin{itemize}
\item The magnetic susceptibility obtained from  the HRG  model \cite{lattice:2014} (dashed curve) confirms the nature of the QCD matter as dia-magnetic at low temperature. Here, the free energy density is considered as the sum over contributions from hadrons and their resonances with masses lighter than one GeV tends to contribute the hadronic interaction in order to assure negative magnetic susceptibility \cite{lattice:2014}.

\item In PLSM, the free energy density, Eq. (\ref{potential}), is divided into three terms. The first one is the pure mesoinc potential which is obtained from the Lagrangian for pure gauge. The second one gives quarks and antiquarks contributions, which apparently have mesonic fluctuations from both quarks and antiquark flavors. The third term represents the interactions of color charges and gluons. This obviously means that two terms contribute to the hadronic fluctuations, while one term contributes to the gluon interactions.

\item At very low temperatures, the slope of magnetic susceptibility [$\chi (T)$] is apparently negative (inside-box in middle panel). This is a signature about QCD matter as dia-magnetic and apparently confirms different lattice QCD simulations. The negative magnetic susceptibility has been obtained within the Parton-Hadron-String Dynamics approach, as well \cite{Cassing:2014}. Switching to high temperature regime, i.e.  restoring the broken chiral symmetry, we observe a transition between dia- and para-magnetic properties. QCD  matter as para-magnetism is very likely at high temperature. The non-interacting MIT bag model \cite{MIT:2008} confirms a phase transition from dia- to para-magnetism. Other QCD-like models can study free quarks coupled to Polyakov loop and give results consistent with the lattice simulations, especially at high temperatures \cite{Orlovsky:2014} .

\item The recent lattice QCD simulations \cite{lattice:2014} (open circle) are estimated by using half-half method in $24^3 \times 32$ lattice (closed triangle) and by using integral method in $28^3 \times 10$ lattice (open triangle). By employing $N_f=2+1$ degrees-of-freedom and by using HISQ/tree action with quark masses $m_l /m_s=0.05$ and temporal dimension $N_\tau=8$, the lattice results are represented by diamonds \cite{lattice:2013b}. The closed circles stand for simulations in isotropic lattice \cite{squeezing:2013}. 

\item The PLSM results seem to confirm that the strongly interacting QCD matter has para-magnetic properties, and its magnetic susceptibility steeply increases towards the deconfinement phase-transition. These conclusions are confirmed in a wide range of temperatures $ 100\leq\,T\,\leq 250$ MeV\cite{Borsanyi:009, Borsanyi:010}.
\end{itemize}

The right-hand panel of Fig. \ref{propes} gives the relative permeability with respect to that of the vacuum compared to recent lattice QCD calculations (open triangles) \cite{lattice:2014} in a wide range of temperatures, at $eB=0.2~$GeV$^2$ and $\mu=0$. There is an obvious quantitative and qualitative agreement between our PLSM calculations and lattice QCD simulations.  From Eq. (\ref{permeability}), one easily realize that $\mu_r$ is very similar to $\chi_B$. The agreement with lattice QCD simulations is thus not surprising. 

 The third part of this work deals with the influences of finite magnetic field on the QCD phase-diagram. This shall be elaborated in section \ref{chiralMF}.


\subsubsection{Influences of finite magnetic field on QCD phase-diagram }
\label{chiralMF}

Here, we introduce other consequences of finite magnetic fields, namely their influences on the QCD phase-diagram. In other words, we analyze how the {\it critical} temperature, even the one corresponding the chemical freezeout, varies with the magnetic field strengths. Two different mechanisms are assumed to play a role. The first one is that the magnetic field improves the phase transition due to its contributions to produce Landau quantizations or levels. With improvement, we mean that the {\it critical} temperature at finite magnetic fields becomes smaller relative to that at vanishing magnetic field. Secondly, the magnetic field contributes to the suppression in the chiral condensates relevant to the restoration of the chiral symmetry breaking. This suppression (and improving) is (are) known as inverse magnetic catalysis and is (are) manifested though CME and magnetic catalysis analysis. 

In determining the {\it critical} temperature and afterwards mapping out the QCD phase-diagram, we implement different methods such as higher-order moments of the quark multiplicity, order parameters, etc. The critical temperature (or baryon chemical potential) can be determined through the intersection of the order parameters, which characterizing the quark-hadron phase transition, the Polyakov-loop fields ($\phi$ and $\phi^*$), with the chiral condensates of light- and strange-quarks, $\sigma_l$ and $\sigma_s$, respectively. The latter is related to the restoration of the broken chiral symmetry. The critical temperature corresponding to chiral restoration of light-quark, $T_c^{\chi _l}$, can be determined from the intersection between $\phi$ and $\sigma_l$, while the critical temperature corresponding to the chiral restoration of strange-quark, $T_c^{\chi _s}$, can be defined from the intersection between $\phi^*$ and  $\sigma_s$.

Alternatively, we might implement the normalized second-order moments of quark multiplicity ($\chi/T^2$) in order to estimate the critical temperature (or $\mu$). In doing this, we analyze the $T$- (or $\mu$-) dependence of $\chi/T^2$ of the system of interest. A peak is conjecture to be located where the critical $T$ (or $\mu$) is reached. 

\begin{figure}[htb]
\centering{
\includegraphics[width=3.9cm,angle=-90]{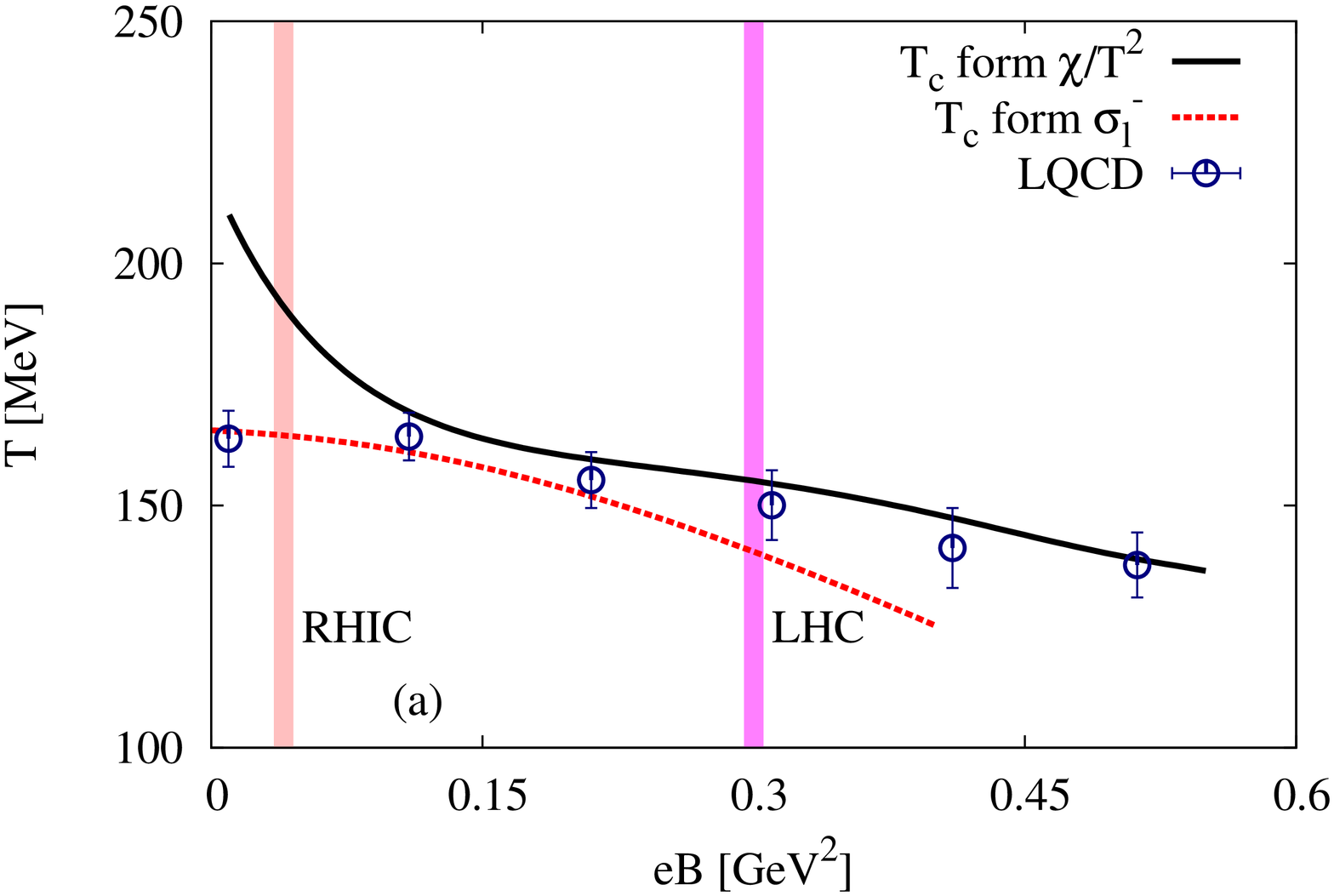}
\includegraphics[width=3.9cm,angle=-90]{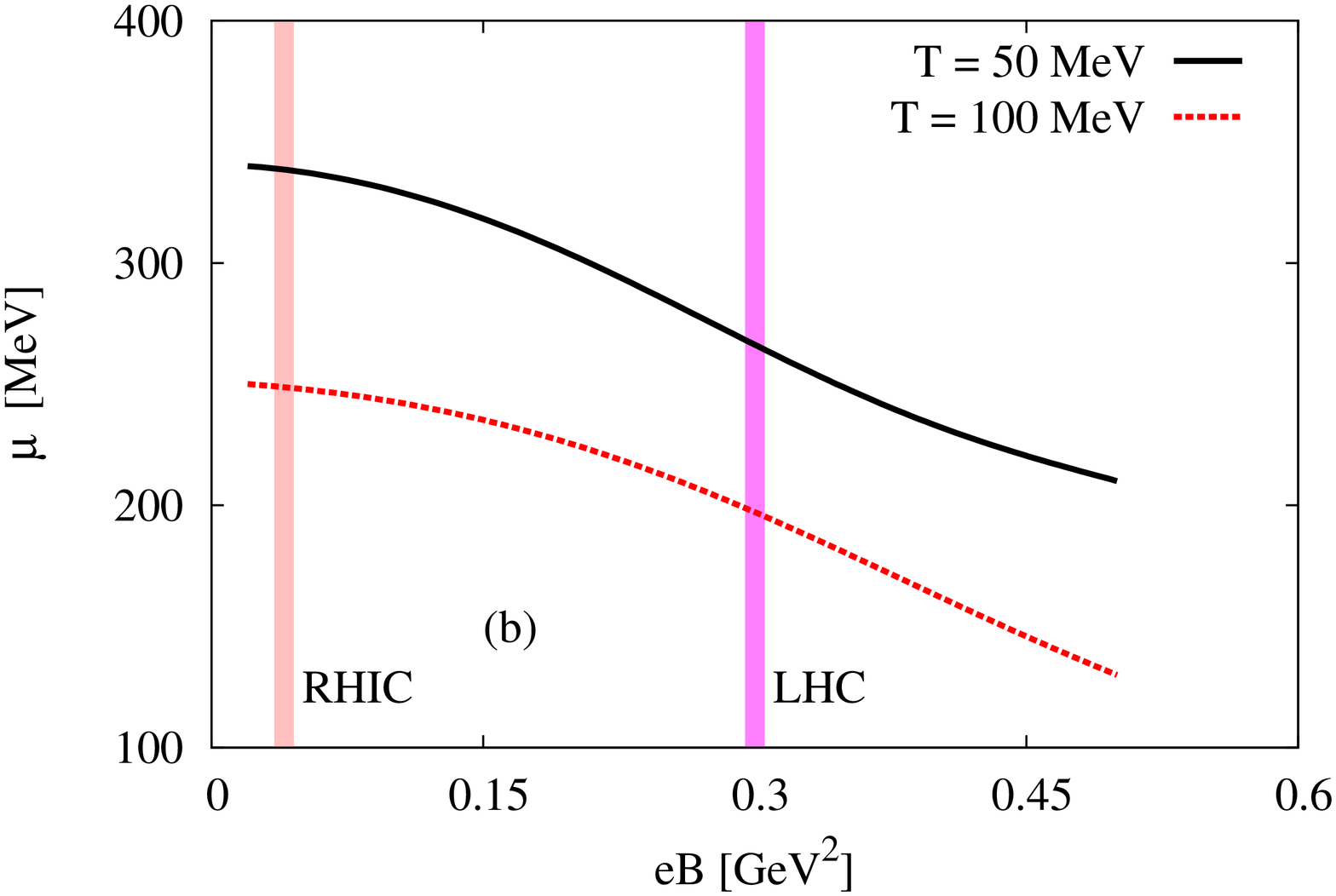}
\includegraphics[width=3.9cm,angle=-90]{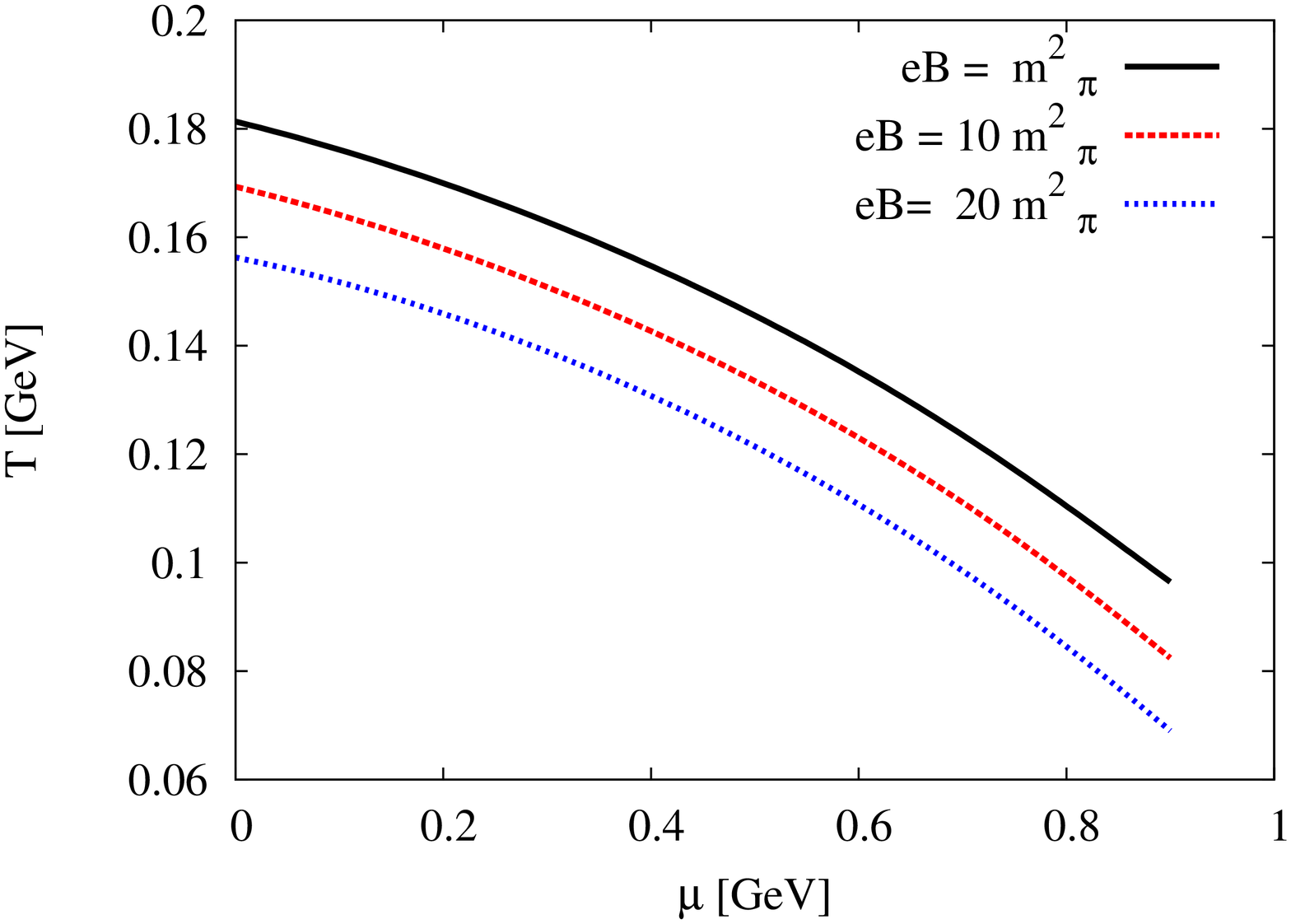}
\caption{(Color online) The chiral phase-diagram relates the critical temperature (left-hand panel) to $eB$ through a method utilizing quark susceptibility $\chi/T^2$ (solid curve) and another one implementing $\sigma_l$ (dashed curve), the critical baryon chemical potential (middle-panel) to $eB$ at $T=50$ (solid) and $T=100~$MeV (dashed curve) and the critical temperature (right-hand panel) to $\mu$ at $eB=m_{\pi}^2$ (solid), $eB=10 m_{\pi}^2$ (dashed), and $eB=20 m_{\pi}^2$ (dotted curve). The vertical bands refer to magnetic field strength estimated at RHIC and LHC energies. \label{fig:eBdepence} 
}}
\end{figure}

\begin{figure}[htb]
\centering{
\includegraphics[width=10.cm,angle=0]{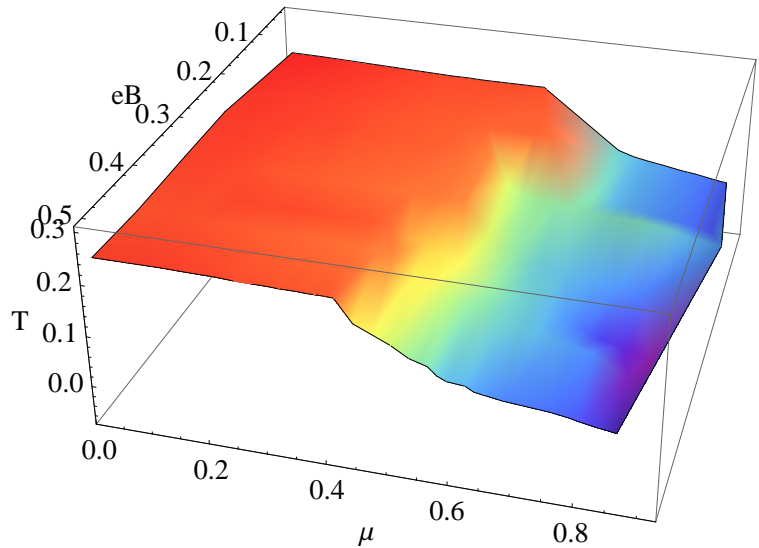}
\caption{(Color online) As in Fig. \ref{fig:eBdepence} but for the chemical freezeout parameters; $T$ vs. $eB$ vs. $\mu$ in GeV units calculated when the freezeout condition $s/T^3=7$ is fulfilled \cite{Tawfik:2014eba}. \label{fig:eBdepence2} 
}}
\end{figure}

 In Fig. \ref{fig:eBdepence}, the QCD phase-diagram in presence of finite magnetic field is computed from the dependence of chiral and/or deconfinement critical temperatures on finite magnetic field.
The possible splitting of  QCD phase-diagram into deconfinement and chiral transitions was introduced and worked out in Refs.  \cite{Mizher;2010,Fraga:split2013,Fraga:split2012,PRCDiab:2016}. The lattice results are given as circles with errorbars \cite{lattice:2014}. The vertical bands refer to the magnetic field strength expected at RHIC  (orders per percent GeV$^2$ or $\sim m_\pi ^2$) and LHC energies (orders per ten GeV$^2$ or $\sim 10-15\, m_\pi ^2$). A small suppression appears in the chiral quark-condensates due to the influence of finite magnetic field. This phenomena is know as {\it inverse} magnetic catalysis.  

In left-hand panel of Fig. \ref{fig:eBdepence}, the solid curve gives the critical temperature estimated from the peaks of the second-order moments of quark multiplicity, the normalized quark susceptibility ($\chi/T^2$). We find that $T$ decreases with increasing the magnetic field strengths (referring to {\it inverse} magnetic catalysis). An excellent agreement is apparently achieved when confronting the dotted curve, which was obtained from the intersection of $\sigma_l$ with $\phi$, to the lattice QCD calculations, especially at low magnetic field; $0\leq\,eB\,$ [GeV$^2$]$\,\leq 0.2$. The solid curve matches well with the lattice results at a wider range of magnetic fields $0.13\leq\,eB\,$ [GeV$^2$]$\,\leq 0.55$. The $\chi/T^2$-method apparently overestimates the lattice calculations at low temperature, while the $\sigma_l$-method slightly underestimates these at high temperature. We conclude that the magnetic field seems to improve the chiral quark-condensates. This depends on the type of contributions to the Landau levels which are introduced to the system. 
 
The middle-panel draws the dependence of resulting $\mu$ on the magnetic field at finite temperatures; $T=50$ (solid) and $100~$MeV (dashed curve). We notice that at constant magnetic field strength as that at RHIC or LHC energy, large $\mu$ can be reached at low temperature, i.e. $\mu$ obviously decreases with increasing $e B$. There is no lattice simulations demonstrating the change in $\mu$ with the magnetic field to compare with. 

The right-hand panel gives $T$-$\mu$ phase-diagram at $eB=m_{\pi}^2$ (solid), $10\, m_{\pi}^2$ (dashed), and $20\, m_{\pi}^2$ (dotted curve). Increasing magnetic field seems to improve, i.e. reduces the critical temperature, the chiral phase-diagram as result of the superstition on the chiral condensates, i.e. the chiral phase transition takes place earlier (at lower temperatures) than the one at $e B=0$.

In a future work, we plan to re-analysis $T$-$\mu$ phase diagram at finite magnetic field. So far, there are various experimental results on chemical and thermal freezeout \cite{Tawfik:2014eba}. The estimation of  freezeout parameters; $T$ and $\mu$, in dependence on heavy-ion centralities or impact parameters would allow us to analysis the influence of the magnetic field, experimentally \cite{Elec:Magnet}.

In Fig. \ref{fig:eBdepence2}, the chemical freezeout condition $s(T, \, eB,\, \mu)/T^3=7$ is implemented \cite{Tawfik:2014eba}. The entropy density is calculated at different temperatures, baryon chemical potentials and magnetic fields. When the entropy density normalized to $T^3$ reaches the value $7$, the values of the freezeout temperature ($T$), the related baryon chemical potential ($\mu$), and the corresponding magnetic field ($e\, B$) are registered. They are three quantities characterizing the chemical freezeout of the system of interest, Fig. \ref{fig:eBdepence2}. For the first time, such a multi-dimensional chemical freezeout boundary illustrating the dependence the {\it ordinary} freezeout diagram ($T-\mu$), which can be directly related to the one analysed from the measurements of various particle ratios \cite{Tawfik:2014eba}, for instance, on finite magnetic field, is presented. It is obvious that, at small $\mu$, the effect of magnetic field is almost negligible.  At higher temperatures, the decrease in $T_c$ around the chiral phase-transition moves to lower temperatures with increasing $e\, B$. Again, this phenomena is known as inverse magnetic catalysis. At very high temperatures, there is a slight increase in $T$ with increasing $e\, B$.  We conclude that increasing $e\, B$ has the effect that the chiral phase-transition takes place earlier (at lower temperatures). It is noteworthy noticing that the shape of $T$-$\mu$ phase-diagram looks different from the one at vanishing $e\, B$ \cite{Tawfik:2014eba}. This shall be analysis in a future work.

%
\section{Conclusion} 
\label{conclusion}

The ultimate goal of the present study is a systematic investigation for temperature and density dependences of the strongly interacting QCD matter from the SU($3$) Polyakov linear-sigma model in presence of finite magnetic field. The introduction of magnetic effects to this model is accompanied by some modifications such as dimensional reduction (changing the phase space as shown in Eq. (\ref{phaseeB}). When the magnetic field is directed along $z-$direction, we can apply the magnetic catalysis property \cite{Shovkovy:2013}, where the dimensions are reduced, $D \rightarrow D-2$, i.e. integral over three-momentum shall be transformed into an integral over one-momentum, e.g. along the direction of the magnetic field, i.e. $z-$direction). Also, the dispersion relation shall be modified and the Landau quantization shall be implemented. For the latter, we use Landau theory for quantized cyclotron orbits of charged particles in external magnetic-field. Consequently, some restrictions are added to the color/electric charges of the quarks. 

By using mean field approximation, we have constructed the PLSM partition function. Then, we have estimated the temperature dependence of the deconfinement  order-parameters ($\phi$ and $\phi^*$) and chiral quark-condensates ($\sigma_l$ and $\sigma_s$) in presence of finite magnetic field. We conclude that, the magnetic field plays an essential role on the QCD phase-transition. The strong magnetic field which is likely generated in heavy-ion collisions, leads to sharp and fast QCD phase-transition. 

The distribution of Landau levels has been studied in order to show how they are occupied at finite magnetic field, temperature and baryon chemical potential. The Landau level occupation, Eq. (\ref{MLL}), varies with the quark electric charge besides $T$ and $\mu$ and is characterized by QCD energy scale.

We have shown that the PLSM in presence of finite magnetic field is in a good agreement with various recent QCD lattice calculations for different magnetic properties such as magnetization, magnetic susceptibility and permeability. The magnetic susceptibility, which can be deduced from the second derivative of PLSM free-energy at finite volume with respect to the magnetic field, is able to highlight the magnetic fluctuations of the strongly interacting QCD matter. It is expected that, at low temperature, the QCD matter creates an induced magnetic field (dia-magnetic material). This has been confirmed by recent lattice simulations. With increasing temperature, the magnetic nature of the QCD matter changes (becomes para-magnetic at high temperatures). 

In addition to the pure mesonic LSM potential which contributes the valence quarks, the Polyakov loops are responsible for integrating the gluon dynamics to the model. The physical mechanism of the magnetic catalysis result from a competition between the valance and sea quarks. When the valence quarks potential has an inefficient effect at high temperature \cite{Tawfik:2014uka}, the contribution of sea quarks will be more than that of the valance quarks.  When the magnetic field is switched on, the temperature dependence of the chiral quark-condensates is remarkable affected, while that of the Polyakov loops fields remains almost unchanged. This implies a suppression of the chiral quark-condensates. To explain this, there are two different mechanisms to propose. The first one suggests that, the magnetic field improves the phase-transition as a result of its contributions to the Landau quantizations. The second one deals with the magnetic field contributes to a suppression in the chiral quark-condensates, which signatures the restoration of the chiral symmetry breaking. This suppression is known as ''{\it inverse magnetic catalysis}'' and defines that, the increase in the magnetic field results in a decrease in the corresponding critical temperature.  In other words, the magnetic field accelerates the phase-transition, i.e. reduces the corresponding critical temperature. Furthermore, we find that the critical temperatures should not be necessarily a universal value. They seem to be depending on quark favors and the magnetic field, as well, which in turn is related to the centrality of the heavy-ion collisions (the impact parameter). 

In two different methods, we have calculated the QCD phase-diagram ($T_c$ vs. $eB$). First from normalized susceptibility and second from the intersection between deconfinement order-parameters and light-quark condensates. The results confirm recent lattice simulations. Furthermore, the dependence of the critical baryon chemical potential on the magnetic field has been determined, as well. We find that increasing magnetic field decreases the critical baryon chemical potential.  This is a known feature of 	the various QCD-like models that, at low temperatures and up to a certain value of $e B$, the critical baryon chemical potential decreases as $e B$ increases \cite{Preis:2011A,Preis:2011B}. Accordingly, we are able to map out $\mu_c$ vs. $eB$ chiral phase-diagram. So far, no lattice calculations are available to compare with. The {\it ordinary} QCD phase-diagram; $T$ vs. $\mu$, is depicted at different magnetic fields. Similar to temperatures, we notice that, increasing magnetic field allows the chiral phase-transitions to take place at lower $\mu$.  

The present paper claims to correct earlier calculations published by one of the authors (AT) \cite{Tawfik:2014hwa}. The main reason for this correction is the zero Landau level. It is conjectured that this is properly taken into consideration in the present work.


\end{document}